\documentclass[12pt]{article}
\usepackage{epsf}

\usepackage{epsfig,graphics}
\usepackage {graphicx}
\usepackage {epsfig}
\usepackage {subfigure}
\usepackage {tabularx} 
\usepackage{rotate}	
\usepackage{slashed}

\usepackage{amsmath}
\usepackage{amsfonts}
\usepackage{amssymb}
\usepackage{graphicx}
\usepackage{cite}
\usepackage{color}
\usepackage{bbm}

\usepackage{fancyhdr}
\usepackage{hyperref}

\newcommand{\op}{\hspace{1pt}}
\newcolumntype{M}[1]{>{\centering\arraybackslash}m{#1}}
\newcolumntype{N}{@{}m{0pt}@{}}

\newcommand{\bmat}{\left(\begin{array}}
\newcommand{\emat}{\end{array}\right)}

\def\yzero{\smash{\hbox{$y\kern-4pt\raise1pt\hbox{${}^\circ$}$}}}

\def\beq{\begin{equation}}
\def\eeq{\end{equation}}
\def\beqa{\begin{eqnarray}}
\def\eeqa{\end{eqnarray}}

\def\-{\hphantom{-}}

\def\s2{\frac{1}{\sqrt2}}

\def\beq{\begin{equation}}
\def\eeq{\end{equation}}
\def\beqa{\begin{eqnarray}}
\def\eeqa{\end{eqnarray}}

\def\IF{\relax{\rm I\kern-.18em F}}
\def\II{\relax{\rm I\kern-.18em I}}
\def\IP{\relax{\rm I\kern-.18em P}}
\def\IC{\relax\hbox{\kern.25em$\inbar\kern-.3em{\rm C}$}}
\def\IR{\relax{\rm I\kern-.18em R}}

\def\cc{{\mathcal C}}

\def\cn{{\mathcal N}}
\def\cam{{\mathcal M}}

\def\cv{{\mathcal V}}

\def\cw{{\mathcal W}}

\def\Dsl{\,\raise.15ex\hbox{/}\mkern-13.5mu D} 

\def\mp{{M_{\text{P}}}}
\def\ms{{M_{\text{s}}}}
\def\mkk{{M_{\text{KK}}}}
\def\mkkt{{M^2_{\text{KK}}}}
\def\mst{{M^2_{\text{s}}}}
\def\mpt{{M^2_{\text{P}}}}
\def\mpf{{M^4_{\text{P}}}}
\def\cco{{\mathfrak{c}}}
\def\nco{{\mathfrak{n}}}
\def\mmod{{M_{\text{mod}}}}
\def\mmodt{{M^2_{\text{mod}}}}
\def\mkky{{M^y_{\text{KK}}}}

\catcode`\@=11   
\newdimen\@rotdimen
\newbox\@rotbox  

\def\@vspec#1{\special{ps:#1}}
\def\@rotstart#1{\@vspec{gsave currentpoint currentpoint translate
   #1 neg exch neg exch translate}}
\def\@rotfinish{\@vspec{currentpoint grestore moveto}}
\def\@rotr#1{\@rotdimen=\ht#1\advance\@rotdimen by\dp#1%
   \hbox to\@rotdimen{\hskip\ht#1\vbox to\wd#1{\@rotstart{90 rotate}%
   \box#1\vss}\hss}\@rotfinish}
\def\@rotl#1{\@rotdimen=\ht#1\advance\@rotdimen by\dp#1%
   \hbox to\@rotdimen{\vbox to\wd#1{\vskip\wd#1\@rotstart{270 rotate}%
   \box#1\vss}\hss}\@rotfinish}%
\def\@rotu#1{\@rotdimen=\ht#1\advance\@rotdimen by\dp#1%
   \hbox to\wd#1{\hskip\wd#1\vbox to\@rotdimen{\vskip\@rotdimen
   \@rotstart{-1 dup scale}\box#1\vss}\hss}\@rotfinish}%
\def\@rotf#1{\hbox to\wd#1{\hskip\wd#1\@rotstart{-1 1 scale}%
   \box#1\hss}\@rotfinish}%
\def\rotate{\@ifnextchar[{\@rotate}{\@rotate[l]}}
\def\@rotate[#1]#2{\setbox\@rotbox=\hbox{#2}\@nameuse{@rot#1}\@rotbox}

\catcode`\@=12

\topmargin
-1.5cm
\textwidth
15.5cm
\textheight
23.5cm
\oddsidemargin
0.7cm
\evensidemargin
0.7cm

\begin{document}

\makeatletter
\@addtoreset{equation}{section}
\makeatother
\renewcommand{\theequation}{\thesection.\arabic{equation}}
\pagestyle{empty}
\vspace{-0.2cm}
\rightline{ IFT-UAM/CSIC-19-159}
\vspace{1.2cm}
\begin{center}


\LARGE{On scale separation in type II AdS flux vacua\\ [13mm]}
  \large{ A. Font$^1$, A. Herr\'aez$^2$ and L.E. Ib\'a\~nez$^2$,
   \\[6mm]}
\small{
$^1$  {\em Facultad de Ciencias, Universidad Central de Venezuela, A.P.20513, \\[-0.3em]
Caracas 1020-A,  Venezuela
}  \\[0pt]
  $^2$ {\em Departamento de F\'{\i}sica Te\'orica
and Instituto de F\'{\i}sica Te\'orica UAM/CSIC,\\[-0.3em]
Universidad Aut\'onoma de Madrid,
Cantoblanco, 28049 Madrid, Spain} \\[5mm]
}
\small{\bf Abstract} \\[6mm]
\end{center}
\begin{center}
\begin{minipage}[h]{15.22cm}
We study the separation of AdS and Kaluza-Klein (KK) scales in type II 4d AdS orientifold vacua.  We first address this problem 
in toroidal/orbifold  type IIA vacua with metric fluxes, corresponding to compactifications in twisted tori, both from the 4d and 10d points of view.
We show how the naive application of the effective 4d theory leads to results which violate the AdS distance conjecture,
in a class of $\cn=1$ supersymmetric models which have a 10d lifting to a compactification on $S^3\times S^3$. 
We show how using KK scales properly modified by the compact metric 
leads to no separation of scales with $\mkkt = \cco |\Lambda|$, with $\cco$ a numerical constant independent of fluxes.  
This applies 
with no need to keep non-leading fluxes fixed.
We also consider a class of IIB models with non-geometric fluxes in which the effective field theory analysis
seems to lead to a naive separation of scales and a violation of the AdS distance conjecture. It has a T-dual which again
may be understood as a 10d type IIA theory compactified on $S^3\times S^3$. In this geometric dual one again observes that
the strong AdS distance conjecture is obeyed with $\mkkt=\cco' |\Lambda|$,  if one takes into account the curvature in the internal space.
These findings seem to suggest that all toroidal/orbifold models with fluxes in this class obey $\mkkt = \cco |\Lambda|$ with
$\cco$ a flux-independent numerical constant.
\end{minipage}
\end{center}
\newpage
\setcounter{page}{1}
\pagestyle{plain}
\renewcommand{\thefootnote}{\arabic{footnote}}
\setcounter{footnote}{0}

\tableofcontents

\section{Introduction}

In the last few years there have been important efforts in trying to ascertain when a
low energy effective field theory may be embedded into a consistent theory of quantum gravity.
Failing to do so locates such a theory in the {\it Swampland} of theories. There is no general simple rule 
to learn when a theory is in the Swampland or not, see \cite{swampland, vafafederico, review} for reviews.
Frequently our knowledge of the Swampland territory
is formulated in terms of conjectures which are then attempted to be tested in string theory, assuming that
the latter is a consistent theory of quantum gravity. In the present paper we will be concerned with two
such conjectures, both applying to  AdS vacua. The first is the {\it \mbox{AdS/KK} scale separation conjecture}  (ASSC) which
states that in any AdS vacua there is no separation between the AdS and the lightest Kaluza-Klein (KK) scales,
as stressed recently \cite{Gautason:2015tig,lpv} and earlier \cite{Duff:1986hr,Douglas:2006es,Tsimpis:2012tu}.
The second is the {\it AdS distance conjecture} (ADC) \cite{lpv}, which states that in
AdS vacua with cosmological constant $\Lambda$, as $\Lambda\rightarrow 0$ there is an infinite tower of states with 
masses (in Planck units)
\beq
m\ \simeq \ |\Lambda|^\gamma \ ,
\eeq 
where $\gamma$ is a positive constant. Thus the limit $|\Lambda|\rightarrow 0$ is not smooth, but rather lies 
at infinite distance. A strong version of this conjecture states  that $\gamma=1/2$ in the supersymmetric case.  
These two conjectures are not unrelated since for $\gamma=1/2$ they are actually equivalent (assuming that the tower predicted by the ADC is the KK tower, as in all the cases analysed in this paper)\footnote{A different {\it AdS/moduli scale separation conjecture} posits that the lightest modulus cannot be 
heavier than the AdS scale \cite{Gautason:2018gln, ralph} and it is fulfilled in all our examples.}.

These conjectures have been tested in many string theory examples. However there is a class of type IIA 4d orientifolds 
\cite{Villadoro:2005cu, DeWolfe:2005uu, cfi, Acharya:2006ne, Palti:2008mg, quirant},
which  appear to violate both  \cite{lpv,Grimm:2019ixq,ralph}. In particular there are certain examples with non-vanishing
Romans mass which seem to yield theories in which the AdS and KK scales can be parametrically separated
\cite{DeWolfe:2005uu, cfi}.  
It has been suggested that these models do not obey the strong ADC because the effective field theory fails to capture the 
backreaction effects of the background \cite{lpv}.

In this paper we revisit this issue and study specifically a class of $\cn=1$ supersymmetric
models which have an explicit uplift  to a 10d theory compactified on $S^3\times S^3$ \cite{AF} (see also
\cite{Acharya:2003ii,Koerber:2007jb,Caviezel:2008ik}).
From the effective field theory point of view these are compactifications with metric fluxes. We find that
using the effective action one obtains a violation of the strong ADC. However, when including 
information from the full $S^3\times S^3$ geometry one obtains that $\mkkt=\cco |V_0|/\mpt$, 
with $\cco$ a numerical constant independent from fluxes, and no scale separation.  
Given that $| \Lambda|=|V_0|/\mpt$, 
this is stronger than the original ADC in the sense that  $\cco$ is flux independent and we are not making the non-leading fluxes small.
We also study the case of a type IIB model with non-geometric fluxes which apparently displays 
AdS/KK scale separation. Although we do not have a geometric 10d interpretation of this model,
it turns out to be mirror to a type IIA model of the above class, which admits a 10d uplift as a
compactification on $S^3\times S^3$. In the dual frame one observes again that there is no scale separation
and  $\mkkt=\cco |V_0|/\mpt$, with $\cco$ flux independent.  
The results in these examples could hint that, after including backreaction effects,
the  class of supersymmetric AdS vacua in \cite{DeWolfe:2005uu, cfi} have all $\mkkt=\cco |V_0|/\mpt$, 
in agreement with the strong ADC. 

Another interesting property that we have observed is that the ratio of the moduli masses to $|\Lambda|^{1/2}$ 
behaves as the ratio $\mkk/|\Lambda|^{1/2}$.  In fact, this also happens in non-supersymmetric vacua found in the 4d effective
theory where our 10d analysis does not directly apply. Nonetheless, in the non-supersymmetric case the 4d results for the masses 
are consistent with no scale separation and the strong ADC.

The rest of this is structured as follows. In sections \ref{s:2a} and \ref{s:2b} we describe the IIA and IIB scenarios
and discuss their main features. Section \ref{s:fin} is devoted to further observations.

\section{A class of type IIA vacua}
\label{s:2a}

In this section we introduce the class of type IIA vacua in which the Swampland conjectures
for different scales will be tested. They are $\cn=1$ supersymmetric solutions of 10-dimensional type IIA 
supergravity with a warped product geometry
\beq
ds_{10}^2 = e^{2A(y)} \hat g_{\mu\nu} dx^\mu dx^\nu + g_{mn} dy^m dy ^n \, ,
\label{met10}
\eeq
where $\hat g_{\mu\nu}$ and $g_{mn}$ are respectively the metrics of $\text{AdS}_4$ and the internal space
$\cam_6$. On top there are background fluxes for the NS-NS 3-form $H$ and the RR $p{\text{-forms}}$, $F_p$, 
$p=0,2,4,6$. In particular, the Romans mass parameter corresponds to the flux of $F_0$.
The complete form of such solutions was obtained in \cite{Lust:2004ig}, extending work of 
\cite{BC}, and rederived in \cite{Grana:2006kf}.
We now briefly summarize the results. The dilaton and the warp factor are constant, related by $\phi=3A$.
The internal space has SU(3) structure, characterized by a real 2-form $J$ and a complex 3-form $\Omega$, 
and only two non-vanishing torsion classes. In this paper we focus in
the nearly K\"ahler case where only the class $\cw_1$ is different from zero. The background
fluxes are determined in terms of $J$, $\Omega$, $A$, plus constants denoted $m$ and $\tilde m$ in \cite{Grana:2006kf}.
The 4-dimensional cosmological constant is determined to be $\Lambda=-3(m^2 + \tilde m^2)$.
The residual constraint from the Bianchi identity for $F_2$ can be fulfilled by adding smeared O6-planes and/or D6-branes.
For certain choices of the parameters it is possible to avoid O6-planes, or even sources altogether \cite{BC, Lust:2004ig}.

In this paper we consider a particular example of nearly K\"ahler $\cam_6$ given by $S^3 \times S^3$. 
In \cite{AF, Caviezel:2008ik} it was shown that compactification on $\text{AdS}_4 \times S^3 \times S^3$ admits an effective 
4d description by including appropriate geometric fluxes in the superpotential. 
Below we will first review the effective approach in $d=4$ and then discuss the lift to $d=10$ following \cite{AF}.

The 4-dimensional setup is that of type IIA toroidal orientifolds. We will mostly
follow the conventions of \cite{cfi}.
We consider a model with moduli consisting of three K\"ahler moduli $T_i$, together with
complex structure moduli separated into the dilaton $S$ and three  $U_i$. The potential for these fields is generated
by RR, NS-NS and geometric fluxes. The fluxes are chosen so that there is a vacuum solution with
$T_i=T$ and $U_i=U$. To simplify we can then restrict from the beginning to fields
$S$, $U$ and $T$.  
According to the standard form of $\cn=1$ supergravity, the scalar potential reads
\beq
\label{fpot}
V= e^K\left\{K^{I\bar J} D_I W \overline{(D_J W)} - 3 |W|^2\right\} \, , 
\eeq
where $K^{I\bar J}$  is the inverse of $K_{I\bar J} = \partial_I\partial_{\bar J} K$, $D_IW = \partial_I W + K_I W$, 
and $I$ runs over ${S,T,U}$. 

The K\"ahler potential is split as $K=K_K + K_Q$, where $K_K$ and $K_Q$ depend on K\"ahler
and complex structure moduli respectively.  
In the large volume regime the K\"ahler piece is given by
\beq
K_K = -\log (8 \cv)= -3\log(T+\bar T) - \log{\cc} \,.
\label{kkpot}
\eeq 
In the second equality we have used that $\cv$ is the volume of the internal manifold  defined in
terms of the K\"ahler 2-form through
\beq
\cv = \frac16 \int_{\cam_6}\hspace*{-4mm} J^3 = \frac{\cc}8 (T+\bar T) ^3 \, .
\label{vdef} 
\eeq
The normalization constant $\cc$ will be specified by consistency with the 10-dimensional analysis.
On the other hand, in the large complex structure limit
\beq
K_Q=4 \phi_4 = - \log (S+\bar S) - 3 \log(U+\bar U) - 2\log{\cc} \, .
\label{kqpot}
\eeq 
Here $\phi_4$ is the 4-dimensional dilaton related to the 10-dimensional one by
$e^{\phi_4} = e^\phi/\sqrt{\cv}$. 

The flux-induced superpotential takes the form
\beq
\frac{W}{\cc}= e_0 + 3 i e T + 3 c T^2 + i M T^3 + i h_0 S - 3 i h U - 3 a S T - 3 b T U \, .
\label{wgen}
\eeq 
The parameters $M$, $c$, $e$ and $e_0$ correspond to fluxes of $F_p$, $p=0,2,4,6$, whereas $h_0$ and $h$
are fluxes of $H$. 
The terms mixing $T$ with $S$ and $U$ are due to the geometric fluxes denoted $a$ and $b$.
Turning on geometric fluxes implies that the internal space is a so-called twisted torus having a basis
of 1-forms $\eta^1, \ldots, \eta^6$, satisfying relations such as 
$d\eta^1 = - a \eta^{56} - b \eta^{23}$, with $\eta^{56}=\eta^5 \wedge \eta^6$ and so on.
The geometric fluxes $a$ and $b$ satisfy the Bianchi identity as checked by taking a further exterior derivative \cite{cfi}.
Consistency of the twisted torus structure requires quantized geometric fluxes \cite{BOOK}. 
The factor of $\cc$ in $W$ arises from the normalization 
$\int_{\cam_6}\hspace*{-2mm} \eta^1 \wedge \ldots \wedge \eta^6=\cc$.
As explained in detail in \cite{cfi}, the fluxes further contribute to tadpoles of the RR 7-form $C_7$ that 
couples to D6-branes and O6-planes. 

It is convenient to work with dimensionless superpotential and K\"ahler potential. The units can be restored by inserting appropriate
factors of the Planck mass $\mp$, which actually appear writing the 4d action in Einstein frame after dimensional reduction.
In particular, in this way the scalar potential picks up a factor of $\mpf$. 
The relation between the string mass $\ms$  and $\mp$ reads
\begin{equation}
\label{stringmass}
\mst=\dfrac{g_s^2 \mpt}{4\pi e^{2 A} \cv}\, .
\end{equation}
Here $g_s=e^{\phi}$ whereas $\cv$ is the volume of the internal manifold.
The warp factor enters because the  metric that appears in the 4d Einstein frame is $\hat g_{\mu\nu}$.
In fact, $e^{2A} \cv$ corresponds to the warped internal volume for constant $A$.
We will be interested in the Kaluza-Klein (KK) mass scale $M_{\text{KK}}$. 
For constant warp factor it can be evaluated as
\begin{equation}
\label{KKmass}
M_{\text{KK}}\sim \dfrac{\ms}{e^{-A} \cv^{1/6}} \sim \dfrac{g_s \mp}{\cv^{2/3}} \, .
\end{equation}
This estimate for $M_{\text{KK}}$ corresponds to the lowest eigenvalue of the Laplacian
in a well defined internal space $\cam_6$.  The largest size of $\cam_6$ is approximated by $\cv^{1/6}$,
but it could be refined.
In the above it is understood that $\cv$, $A$, and $g_s$, are all evaluated at the moduli vevs.

AdS supersymmetric minima are straightforward to obtain by solving $D_IW=0$. 
The full results can be found in \cite{cfi}.
We will shortly analyze two particular examples with $M=0$ and $M\not=0$.
Our main purpose will be to study the various scales, namely the KK mass, the cosmological constant
and the moduli masses. To this end we need to connect with the 10d description of the vacua in order
to determine the value of the constant $\cc$ in the normalization of the internal volume.

In \cite{AF} it was shown that the AdS solutions found in the 4d effective approach fully conform
to the results predicted by the general 10d analysis.
To begin, the pair $(J,\Omega)$ of the twisted torus coincides with the nearly K\"ahler SU(3) structure
of $S^3\times S^3$. In particular, the K\"ahler form can be written as
\beq
J=\frac{t}{2\sqrt{ab^3}} \sum_{i=1}^3 \xi^i \wedge \hat \xi^i \, ,
\label{Js3s3}
\eeq
where $\xi^i$ and  $\hat\xi^i$ are left-invariant forms of SU(2). Next, taking into account the moduli vevs found in 4d,
the background fluxes can be expressed in terms of $(J,\Omega)$ precisely as dictated by the 10d analysis. The parameters
that determine the cosmological constant in the 10d formulation are given by $m=\frac{M}{5} e^{4A}$ and 
$\tilde m = \frac{3(c- M \text{Im}\, T)}{t} e^{4A}$, with $A=\frac{\phi}3$.

The explicit expression of the  K\"ahler form $J$ allows to compute the volume of the internal manifold. 
It follows that $\cv=\cc t^3$, where
\beq
\cc = \frac{(4\pi)^4}{(4ab^3)^{3/2}} \, .
\label{valuec}
\eeq
The dependence of $\cc$ on the geometric fluxes wil be crucial in the ensuing discussion. We remark that this dependence 
can only be obtained from the full 10d solution, including the details of the backreacted internal geometry. 
Then, our philosophy in the following will be to compare the results obtained from the naive application of  the purely 4d EFT 
(i.e. without including the factors of $\cc$) with those in which the full 10d theory is used to determine the internal geometry as in \cite{AF}, and the non-trivial dependence of $\cc$ on the geometric fluxes is included.  

\subsection{Example 1,  $M=0$ }
\label{ss:ex1}

We first consider an example without Romans mass, namely without flux for $F_0$.
To further simplify we set to zero the NS-NS fluxes $h_0$ and $h$, as well as the RR 4-form flux $e$.
In this case the K\"ahler axion and only one combination of the remaining axions are fixed as
\beq
{\text{Im}}\, T=0, \qquad a\, {\text{Im}}\, S + b\, {\text{Im}}\, U=0 \, .
\label{ax1}
\eeq
The saxions $s={\text{Re}}\, S$, $u={\text{Re}}\, U$ and $t={\text{Re}}\, T$ are all stabilized at
values
\beq
s=\frac{2 c}{a} t \, ,  \quad u=\frac{6 c}{b} t  \, , \quad t^2 = \frac{e_0}{9 c} \, .
\label{sax1}
\eeq
Without loss of generality we can choose $e_0 >0$ so that necessarily $c, a, b > 0$.
Since the flux $e_0$ is not constrained by tadpole cancellation, it can be taken large to have
large vevs for all saxions. On the other hand, to stay in perturbative regime with both couplings $e^{\phi_4}$ and $e^{\phi}$
small, it is necessary to take $c$ large. However, this coupling is constrained by cancellation of tadpoles that receive
contributions proportional to $ca$ and $cb$ \cite{cfi}.

The cosmological constant is determined by the value of the potential at the minimum, denoted $V_0$.
Inserting the above vevs for the moduli yields
\beq
\frac{V_0}{\mpf} = -\frac{a b^3}{128 \op \cc c^2 t^3} = 
- \frac{27 a b^3}{128 \op \cc c^{1/2} e_0^{3/2}}\, .
\label{V01}
\eeq
We remark that the cosmological constant calculated in the 10d analysis matches the above result once the
Planck units are restored \cite{AF}.

The masses of the canonically normalized moduli are derived by diagonalizing the matrix
$\frac12 K^{ij} \partial_i \partial_j V$ evaluated at the minimum. They are found to be proportional
to $|V_0|/\mpt$. Specifically
\beq
M^2_{\text{mod}} = \left\{6, \frac{22}{27}, -\frac{2}{3}, \frac{10}{3}, -\frac{8}{27},0\right\} \frac{|V_0|}{\mpt}  \, .
\label{mmod1}
\eeq
The eigenvectors corresponding to the first (last) three entries are combinations of saxions (axions).
As expected for a supersymmetric minimum,
the negative eigenvalues are above the Breitenlohner-Freedman bound $m^2 \ge -\frac34 |V_0|/\mpt$ \cite{BFbound}. 
The zero eigenvalue correlates with the combination of axions that remains unfixed.
The behavior $M^2_{\text{mod}} \sim |V_0|/\mpt$ is actually expected for generic tree-level flux vacua \cite{ralph}.

Let us now study the KK scale. Substituting the vevs in \eqref{KKmass} gives
\beq
\frac{M_{\text{KK}}}{\mp} = \frac{(a b^3)^{1/4}}{\cc^{2/3} c^{1/4} e_0^{3/4}} \, .
\label{KK1}
\eeq
Here and below we will omit purely numerical factors to avoid cluttering.
The relevant ratio to study both the ASSC and the strong ADC conjectures is $\mp \mkk/|V_0|^{1/2}$, which reads
\begin{equation}
\dfrac{\mp \, \mkk}{|V_0|^{1/2}}=\frac{1}{(ab^3)^{1/4} \cc^{1/6}}.
\label{ratioAdS1}
\end{equation}
It is then evident that the naive 4d EFT approach could give misleading results.
Without including the effect of $\cc$, effectively setting $\cc=1$ above, would lead to a flux
dependent ratio between the KK and the cosmological constant scale. However, the value of $\cc$
derived in the 10d formulation, cf.~\eqref{valuec}, is such that the dependence of the ratio \eqref{ratioAdS1}
on the geometric fluxes $a$ and $b$ drops out altogether. Thus, in the end we obtain the exact behavior predicted
by the strong ADC not only in the limit $e_0 \to \infty$, with other fluxes fixed, but for all values of the fluxes.

Let us observe that this first model serves to illustrate how the 10d solution changes the purely 4d picture in an 
interesting way, providing an exact flux cancellation in the relevant ratio of scales. 
However, we must remark, as noticed also in \cite{ralph},  
that the 4d EFT still did not have parametric scale separation nor gave a controllable counterexample to the strong ADC per se. 
The reason is that the two interesting limits in the fluxes in which this could occur, namely 
$a,b\rightarrow 0$ or $a,b\rightarrow \infty$ cannot be taken arbitrarily. 
The former cannot be explored because this family of  solutions requires non-vanishing values of $a$ and $b$ and 
since they are quantized we cannot make them go continuously to zero.
In the latter case, since the fluxes $a$ and $b$ enter the tadpole cancellation conditions, making them arbitrarily large would 
imply an unbounded number of D6-branes wrapping internal 3-cycles, whose backreaction could bring the EFT out of control.
In section \ref{ss:ex2} we will discuss an example in which there is no constraint from tadpole cancellation and the fluxes 
$a,b$ can take arbitrarily large values.

As seen from \eqref{mmod1}, in the 4d EFT the moduli masses satisfy $M_{\text{mod}}=\nco |V_0|^{1/2}/\mp$, 
with $\nco$ an order one (flux independent) constant. 
Such relation is actually valid without including the backreation of the geometric fluxes.
We have further seen that the KK scale computed incorporating 10d effects has the same behavior as
$M_{\text{mod}}$ with respect to the
cosmological constant, in agreement with the strong ADC. 
As this happens in other examples of AdS vacua, 
it could be an indication, assuming that the strong ADC holds, that the ratio of the true KK scale to the cosmological constant
can be estimated by the smallest ratio $\mmod/|V_0|^{1/2}$, even when the 10d lift is not known in detail \cite{ralph}. 

The example in this section, with equal configuration of fluxes, also admits classically stable non-supersymmetric vacua.
The 4d EFT results for the moduli and KK masses are qualitatively the same as in the supersymmetric vacuum.
In Planck units $M_{\text{mod}}\sim |V_0|^{1/2}$ and the naive KK mass would match $|V_0|^{1/2}$ upon
including the same volume correction as in the supersymmetric case.
It would be interesting to see if there is a 10d lift satisfying the general conditions spelled out in \cite{Lust:2008zd}.
A 10d solution would presumably lead to a result for $\mkk$ compatible with  $M_{\text{mod}}$ up to numerical factors.

\subsection{Example 2, $M\not= 0$ }
\label{ss:ex2}

We now turn to an example with $F_0$ flux and in which the geometric fluxes are not constrained by
tadpole cancellation.
To make our point it suffices to consider particular values for other fluxes. Concretely we take
$e_0=0$, $e=0$, $h_0=3 c a/M$ and $h=-c b/M$. 
This choice of NS fluxes is such that RR tadpoles due to the fluxes vanish altogether.
Equivalently, the 10d Bianchi identity for $F_2$ is satisfied without having to add 
smeared sources \cite{BC, Lust:2004ig, AF}. 
The upshot is that the independent fluxes $c$, $M$, $a$ and $b$ are not constrained. 

Solving $D_IW=0$ the axions are found to be
\beq
{\text{Im}}\, T=(\lambda+1)\frac{c}{M}, \qquad 
a\, {\text{Im}}\, S + b\, {\text{Im}}\, U=(8\lambda^2+1)\frac{c}{M} \, ,
\label{ax2}
\eeq
where $\lambda=(80)^{-1/3}$. The saxions are fixed as
\beq
s=-\frac{2 c}{a}\lambda  t \, ,  \quad u=-\frac{6 c}{b}\lambda  t  \, , \quad t^2 = \frac{15 c^2 \lambda^2}{M^2} \, .
\label{sax2}
\eeq
Necessarily $c a < 0$ and $c b < 0$. By taking large $c$ it is possible to attain large vevs for all saxions,
as well as small couplings $e^{\phi_4}$ and $e^{\phi}$.

The cosmological constant is now given by
\beq
\frac{V_0}{\mpf} = -\frac{a b^3}{120 \op \cc \lambda^2 c^2 t^3} = 
- \frac{a b^3 M^3}{120 \sqrt{15} \op \cc \lambda^5 c^5}\, .
\label{V02}
\eeq
The masses of the canonically normalized moduli have the expected relation to the cosmological constant. We find
\beq
M^2_{\text{mod}} = \left\{\frac19(47 + \sqrt{159}), \frac19(47 - \sqrt{159}) ,  \frac13(4 + \sqrt{6}),
\frac13(4 - \sqrt{6}), -\frac{2}{3},
0\right\} \frac{|V_0|}{\mpt}  \, .
\label{mmod2}
\eeq
The negative eigenvalue is above the Breitenlohner-Freedman bound and the corresponding eigenstate is a combination
of $s$ and $u$. The zero eigenvalue is due to the unstabilized combination of axions. Other eigenvectors mix saxions and axions.

The KK mass is determined inserting the vevs of the saxions in \eqref{KKmass}. We obtain
\beq
\frac{M_{\text{KK}}}{\mp} 
=\dfrac{(ab^3)^{1/4}}{\mathcal{C}^{2/3}  \lambda \op c \op t^{3/2}}
= \frac{(a b^3)^{1/4} M^{3/2}}{\lambda^{5/2} \op \cc^{2/3}\op c^{5/2}} \, .
\label{KK2}
\eeq
The ratio to the cosmological constant is then
\begin{equation}
\dfrac{\mp \, \mkk}{|V_0|^{1/2}}
=\frac{1}{(ab^{3})^{1/4} \mathcal{C}^{1/6}}.
\label{ratioAdS2}
\end{equation}
As in the previous case, without incorporating the 10d dependence of $\cc$ on the geometric fluxes would yield a flux-dependent ratio. Moreover, in this model there is a particularly interesting limit that would seem to violate the strong ADC as we now discuss. 

To define the limit correctly, we introduce the parameter $\tau$ which will be sent to infinity. We take 
\begin{equation}
a\sim \tau \, \qquad b\sim \tau\, , \qquad c \sim \tau^{\alpha} \, , 
\end{equation}
where $\alpha$ is a positive constant to be fixed momentarily. Note that the geometric flux dependent 
constant is now  parametrically different from the naive 4d expectation since it goes as $\cc\sim \tau^{-6}$. 
This limit is particularly interesting because it implies
\begin{equation}
  V_0 \sim - \dfrac{\tau^4}{\tau^{5\alpha} \mathcal{C}}, \qquad \dfrac{\mp \, \mkk}{|V_0|^{1/2}}
\sim \frac{1}{\tau \op \mathcal{C}^{1/6}}.
\label{limit2}
\end{equation}
Clearly $V_0\to 0$ in the $\tau\to \infty$ limit, both in the naive 4d picture and in the full 10d solution, provided $\alpha>2$. 
It can be checked that the limit leads to large volume and small couplings $e^{\phi_4}$ and $e^\phi$.
Now, from the purely 4d EFT point of view one could erroneously claim that \eqref{limit2} embodies
a counterexample to the strong ADC, since in the $V_0\to 0$ limit we observe a parametric deviation from 
$\mp \mkk \sim |V_0|^{1/2}$. Once again, this seems to be an artifact of the 4d calculation of the KK scale.
Taking into account the full 10d geometry through the proper value of $\cc$,  the dependence on the geometric fluxes of the 
ratio \eqref{ratioAdS2} identically drops out and there is no parametric separation between the two scales. 
Finally, note that even though the limit $\tau\to \infty$
seems dangerous from the purely 4d EFT, it does not exhibit what is usually referred to as scale separation since ignoring $\cc$
the KK scale becomes arbitrarily lower than $|V_0|^{1/2}$ (instead of higher), thereby bringing the EFT out of control. 
Still, this shows a limit in which the full 10d solution crucially modifies the naive expectations from the lower dimensional 
EFT in a way consistent with the strong ADC. 
Moreover, once again, the relation between the corrected KK scale and $|V_0|^{1/2}$ happens to be captured by 
the mass of the moduli, since even in the naive 4d picture we have $\mp M_\text{mod}=\nco |V_0|^{1/2}$.

\section{A class of type IIB vacua}
\label{s:2b}

In this section we look into a class of supersymmetric type IIB vacua which are dual to the ones studied in the previous section.
 As we will see, from the naive approximation of the KK scale available in the 4d EFT, these vacua seem 
 to exhibit both scale separation and a behavior that contradicts the strong ADC. In these cases, since they include non-geometric 
 fluxes we do not have a full 10d solution which we can use to compute the corrected KK scale and check whether it is modified 
 in such a way that scale separation is not achieved and the strong ADC fulfilled. Our strategy for these vacua is then to use their 
 type IIA duals, whose 10d solution geometry we explained in the previous section, and estimate the corrected KK scale 
 by careful identification on the IIA side.

The non-geometric IIB theory is obtained by performing three T-dualities along the three $x$ directions of the twisted 
torus in IIA \cite{acfi}. In the  4d $\mathcal{N}=1$ theory,  we again restrict ourselves to the case $T_i=T$ and $U_i=U$. 
The scalar potential is calculated by means of the standard eq.~\eqref{fpot}. The K\"ahler potential takes the form
\begin{equation}
K=- \log (S+\bar S) - 3 \log(U+\bar U) - 3 \log (T+\bar T)\, .
\end{equation}
The superpotential is derived by replacing $T \leftrightarrow U$, $S \to S$ in \eqref{wgen}. Thus
\begin{equation}
W=e_0 + 3 i e U + 3 c U^2 + i M U^3 + i h_0 S - 3 i h T - 3 a S U - 3 b T U \, .
\end{equation}
The coefficients are labeled as before but now $h$ and $b$ have a different interpretation in terms of non-geometric fluxes 
in type IIB.  We have not included a possible flux-dependent normalization constant, but will keep in mind that it might
be present.

For completeness we recall the relation between the Planck and the string masses in IIB compactifications, namely 
\begin{equation}
\mst=\dfrac{g_s^{1/2} \mpt}{2 \pi \mathcal{V}},
\end{equation}
where $g_s=e^{\phi}=s^{-1}$ and $\mathcal{V}=t^{3/2}$.
At the level of the EFT, the KK scale is approximated by
\begin{equation}
\mkk \sim \frac{\ms}{V_6^{1/6}}\sim \frac{\mp}{\mathcal{V}^{2/3}},
\label{mkkIIB}
\end{equation}
where we used that in IIB, $V_6 = e^{3\phi/2} \cv$. 
In the above we have neglected warp factors.
Let us remark that at this point all these relations refer to the compact manifold consisting of a flat toroidal orientifold 
with (non-geometric) fluxes, which is the 4d EFT framework. In this case with non-geometric fluxes the 10d picture
is not even in terms of standard spacetime.

\subsection{Example 3, $M=0$}

To simplify the discussion, we again turn off the fluxes $h_0$, $h$ and $e$, and recall that the flux $e_0$ is not 
constrained by any tadpole. As expected by  consistency with the IIA dual, there exists a supersymmetric vacuum, see also 
\cite{Blumenhagen:2015kja}.
The axions are fixed at values
\begin{equation}
{\text{Im}}\, U=0, \qquad a\, {\text{Im}}\, S + b\, {\text{Im}}\, T=0 \, ,
\label{ax3}
\end{equation}
whereas the saxions are stabilized according to
\begin{equation}
s=\frac{2 c}{a} u \, ,  \quad t=\frac{6 c}{b} u  \, , \quad u^2 = \frac{e_0}{9 c} \, .
\label{sax3}
\end{equation}
The value of the potential at the minimum is given by 
\begin{equation}
\frac{V_0}{M_P^4} = -\frac{a b^3}{128 \op c^2 u^3} = 
- \frac{27 a b^3}{128 \op c^{1/2} e_0^{3/2}}\, ,
\label{V03}
\end{equation}
which coincides with eq. \eqref{V01} after changing $t\rightarrow u$ and setting $\cc=1$. 
The masses of the moduli are given by \eqref{mmod1}.
Using \eqref{mkkIIB},  the EFT approximation of the KK scale yields
\begin{equation}
\dfrac{\mkk}{\mp} \sim \frac{1}{t}
\sim \frac{ 1}{(2 \op c \op e_0)^{1/2}}\, .
\label{kk3wrong}
\end{equation}
The ratio of $\mkk$  to the cosmological constant scale takes the form
\begin{equation}
\dfrac{\mp \mkk}{|V_0|^{1/2}}
\sim \frac{e_0^{1/4}}{c^{1/4} (ab)^{1/2}},
\label{ratio3wrong}
\end{equation}
which is in agreement with \cite{Blumenhagen:2015kja}. 

In this example, taking $e_0$ to be large implies a large complex structure point in which the string coupling is small, 
so the EFT stays within its regime of validity. Moreover, upon taking the limit $e_0\rightarrow \infty $,  
scale separation would hold since the KK mass becomes arbitrarily larger than the cosmological constant scale.
In addition, since in this limit $V_0\rightarrow 0$, the strong ADC would also be violated. 
We claim that this is again an artifact of the wrong approximation for $\mkk$.
The supporting arguments rely on the T-duality with the type IIA model of section \ref{ss:ex1}, which has a clear 10d 
interpretation as explained before. 

As already mentioned, to go from IIA to IIB, 
one performs three T-dualities along  the three $x$ directions of the twisted torus.
Upon this transformation, the sizes associated to the three $y$ directions remain untouched, so that 
the KK scale asociated to the $y$ directions in the IIA side maps to the KK scale along the same directions in the IIB dual. 
Knowing the correct KK scale asociated to the $y$ direction in the IIB side, we can in fact justify that it will be 
the dominant one. The reason is that the large complex structure limit corresponds (for rectangular tori)  precisely to 
having typical length along $y$, much larger than the one along $x$, say $R_y \gg R_x$. 
Hence, the lowest KK scale in the IIB side will be given by just dualizing the IIA KK scale associated to $R_y$. 
In both IIA and IIB we can define $\mkky\sim \ms/R_y$. The dependence on the moduli in the two theories
is different but straightforward to derive\footnote{See e.g. \cite{FHI} for a full correspondence between the KK and winding 
scales in toroidal compactifications of type IIA and type IIB.}.

In IIB we just find 
\begin{equation}
\frac{\mkky}{\mp} \sim \frac{1 }{t \op u^{1/2}} \sim \frac{b }{c^{1/4} \op e_0^{3/4}} \, .
\label{kk3y}
\end{equation}
This yields the ratio
\begin{equation}
\frac{\mp \mkky}{|V_0|^{1/2}}\sim \frac{1}{(ab)^{1/2}},
\label{ratio3y}
\end{equation}
in which the dependence on $e_0$ and $c$ has canceled so that the parametric scale separation, as well as the parametric 
violation of the strong ADC disappear. There is still the dependence on the fluxes $a,b$. 
Now, in IIA we can just read off the result for$\mkky/\mp$ from \eqref{KK1} by setting $a=b$. 
This gives a result that coincides with \eqref{kk3y} except for a factor of $\cc^{2/3}$ in the denominator.
Invoking T-duality we can say that this factor is missing in IIB and should be included. Applying the same reasoning to 
$V_0$, the flux dependence in the ratio will cancel. This was not done before because we were only using 
the 4d EFT, which did not predict the factors of $\cc$. Note however that including these factors would not alter the 
previous discusion involving the $e_0\rightarrow \infty$ limit, since $\cc$ is independent of $e_0$. 

In the end we can distinguish three levels of refinement. In the first one, when the KK scale is approximated by \eqref{kk3wrong}, 
we seem to violate both the ASSC and the strong ADC parametrically. After realizing that the isotropic approximation is not justified 
and  using the dominant length scale, $R_y$, to calculate the KK mass as in \eqref{kk3y}, we observe that the parametric 
violation of both conjectures disappears, but still there is some dependence in the fluxes $a$ and $b$. 
Finally, if we take into account the full 10d geometry of the dual and dualize the KK scale along the directions 
that remain untouched, as well as $V_0$, we find a flux independent relation $\mp \mkk\sim |V_0|^{1/2}$.

\section{Final Comments}
\label{s:fin}

The purpose of this paper has been  to test  the ASSC and ADC within the context of 4d toroidal orbifold  
orientifolds in type II string theory.  Concerning the ASSC, it was already known from many examples 
that, at the level of known 10d compactifications,  such a separation does not exist. Here we have focused however on the 
effective 4d effective action description of $\text{AdS}_4$  type II vacua, in which some examples seem to violate 
the conjecture. We have seen that models where the ASSC conjecture seems to be violated at the 4d effective theory 
level, do not contradict it once one takes into account the backreaction within a complete 10d formulation.  
The same happens with the related ADC, which claims a behavior $\mkk \propto  |V_0|^\gamma$  
as $V_0 \rightarrow 0$, with $\gamma$  positive and equal to 1/2 in the supersymmetric case.  
Using the effective field theory approach, a violation is observed in a number of examples. However, after considering the
backreaction of the geometry when a 10d uplift exists, no violation appears. 

One interesting property that we observe, at least in this class of models, is that we actually find 
$\mkkt=\cco |V_0|/\mpt$ with $\cco$ a flux-independent constant. This goes beyond the strong ADC in the 
sense that there is no need to set the non-leading fluxes of the compactification to 
small values, it is valid for any value of the fluxes. This behavior is universal in this class of models and the only model 
dependence is in the numerical constant $\cco$. 
In this sense the behavior is like the one of the Breitenlohner-Freedman bound, which is also universal \cite{BFbound}. 
Another interesting property, already noticed in \cite{ralph}, is that with respect to the cosmological constant,
the moduli masses behave exactly like $\mkk$, up to numerical factors. 
Thus, in AdS vacua, the ratio $\mmodt/|\Lambda|$ determined from the effective potential gives us automatically the 
ratio $\mkkt/|\Lambda|$ in the theory, without the need of an explicit computation. 
It would be interesting to understand these points further. 

Although we focused on $\cn=1$ supersymmetric models with a 10d lift, we also considered
non-supersymmetric vacua in the 4d EFT. In this latter case we found that the mass scales also 
follow the pattern  $\mkk \sim \mmod \sim |V_0|^{1/2}$, so that there is no scale separation and the
strong ADC is satisfied.

There are models like the specific $M\not= 0$ examples  with no geometric fluxes in \cite{DeWolfe:2005uu,cfi} which 
seem to violate both conjectures if analyzed from the 4d point of view. 
These examples admit a 10d extension with smeared sources, 
but a solution with localised O6-planes is not guaranteed \cite{Acharya:2006ne}.
In any case, the evidence found for the examples considered here 
seem to suggest that in that case also the mass of the moduli captures the effective KK scale, and that again the ASSC 
and strong ADC hold.

\bigskip

\centerline{\bf \large Acknowledgments}

\bigskip

\noindent We would like to thank R. Blumenhagen, M. Brinkmann,  F. Marchesano, M. Montero, E. Palti, A. Uranga,  T. Van Riet and I. Valenzuela for useful discussions. 
This work is supported by the Spanish Research Agency (Agencia Estatal de Investigacion) through the grant IFT 
Centro de Excelencia Severo Ochoa SEV-2016-0597, and by the grants 
FPA2015-65480-P and PGC2018-095976-B-C21 from MCIU/AEI/FEDER, UE. 
A.H. is supported by the Spanish FPU Grant No. FPU15/05012 and would like to thank the Max Planck Institute for Physics in Munich for their hospitality during completion of this work. 
A.F. thanks the IFT UAM-CSIC for hospitality and support during completion of this work.

\end{document}